\documentclass[conference, a4paper]{IEEEtran}
\usepackage{times}

\usepackage{graphicx}
\usepackage{color}
\usepackage{placeins}
\usepackage{float}
\usepackage{hyperref}
\usepackage{tabularx,colortbl}
\usepackage{bm}
\usepackage{amsmath}
\usepackage{amssymb}
\usepackage{mathrsfs}

\def\v{\mathbf}
\def\mathscr{\mathcal}
\begin{document}

\title{Sum-Frequency and Second-Harmonic Generation from Plasmonic Nonlinear Nanoantennas}

\author{\IEEEauthorblockN{%
    Xiaoyan~Y.Z.~Xiong\IEEEauthorrefmark{1},
    Li~Jun~Jiang\IEEEauthorrefmark{1},
    Wei~E.I.~Sha\IEEEauthorrefmark{1},
    Yat Hei Lo\IEEEauthorrefmark{1}, and
    Weng Cho Chew\IEEEauthorrefmark{2}}
  \medskip
  \IEEEauthorblockA{\IEEEauthorrefmark{1}Dept. of Electrical and Electronic Engineering, The University of Hong Kong, Hong Kong\\
    e-mails: xyxiong@eee.hku.hk, jianglj@hku.hk, shawei@eee.hku.hk, yathei36@gmail.com}
  \IEEEauthorblockA{\IEEEauthorrefmark{2}%
    Dept. of Electrical and Computer Engineering, University of Illinois at Urbana-Champaign, Urbana, USA\\
    e-mail: w-chew@uiuc.edu}
    }

\maketitle

\begin{abstract}
Plasmonic nanostructures that support surface plasmon (SP) resonance potentially provide a route for the development of nanoengineered nonlinear optical devices. In this work, second-order nonlinear light scattering, specifically sum-frequency generation (SFG) and second-harmonic generation (SHG), from plasmonic nanoantennas is modeled by the boundary element method (BEM). Far-field scattering patterns are compared with the results calculated by the Mie theory to validate the accuracy of the developed nonlinear solver. The SFG from a multi-resonant nanoantenna (MR-NA) and the SHG from a particle-in-cavity nanoantenna (PIC-NA) are analyzed by using the developed method. Enhancements of the scattering signals due to double-resonance of the MR-NA and gap plasmonic mode of the PIC-NA are observed. Unidirectional nonlinear radiation for the PIC-NA is realized. Moreover, its emission direction can be controlled by the location of the nanosphere. This work provides new theoretical tools and design guidelines for plasmonic nonlinear nanoantennas.
\end{abstract}


\section{Introduction}
Plasmonic nanoantennas made from nanostructured metals have attracted significant attention in nonlinear optics due to their unique properties [1]. One such property is their ability to concentrate light in nanoscale volumes and subsequently boost the intensity of local fields near particle surfaces due to surface plasmon resonance (SPR) [2]. The SPR-enhanced near-fields allow weak nonlinear processes, which depend superlinearly on the local fields, to be significantly amplified giving rise to a promising research area called nonlinear plasmonics. The second-order nonlinear processes, e.g., sum-frequency generation (SFG) and second-harmonic generation (SHG), are significantly dependent on the symmetry of both the material being used and the structure being studied [3]. They are forbidden in the bulk of centrosymmetric media under the electric dipole approximation. However, the breaking of inversion symmetry at surfaces results in surface nonlinear scattering [4-5]. The combination of nonlinear surface sensitive and strong near-field enhancement associated with SPR provides a unique tool for ultrasensitive shape characterization [6-7], super-resolution imaging, sensing and microscopy [8-10], on-chip optical frequency conversion, switching and modulation [11], etc.

Efficient nonlinear scattering requires the presence of strong nonlinear polarization sources at the surface of nanostructures, as well as efficient scattering of the signal into the far-field. Analogous to classical antenna, the objective of nonlinear nanoantenna design is the optimization and control of the spatial distribution of scattered light. Several strategies have been applied to enhance the scattered nonlinear signals, including engineering near-field coupled nanoparticle clusters associated with Fano resonances [12-13], enhancing
the electric fields using nanogaps [14], etc. However, tailoring the radiation pattern of nonlinear nanoantennas has great challenges. First, nonlinear radiation exhibits complex multipolar interactions [15-16]. Second, currently few tools are available for efficient and rigorous analyses of surface nonlinear scattering processes in complex structure. The volume discretization based full-wave methods, e.g., finite-difference time-domain (FDTD) and finite element method (FEM), are inefficient and inaccurate for solving surface nonlinear problems. Third, the relation between plasmon resonances, particle geometry, and associated local-field distributions is very complicated [17]. Physical principles and design rules for nonlinear nanoantennas have not been explored yet.

In this work, a numerical solution based on the boundary element method (BEM) is proposed for the second-order surface nonlinear scattering problems. The proposed method is efficient with a surface discretization; and can employ experimentally tabulated material parameters directly. The developed solver is utilized to systematically analyze the SFG from a multi-resonant nanoantenna (MR-NA) and SHG from a novel particle-in-cavity nanoantenna (PIC-NA) with strong SPR. The remaining of this article is organized as follows. In the next section, we describe the BEM method used in our study. Sec. 3 contains the main results. We first validate the accuracy of the developed solver by comparing the far-field scattering patterns with Mie theory results. Then, the SFG from MR-NA and SHG from PIC-NA are analyzed. Finally, in the last section, we summarize the main conclusions of our work.


\section{Methods}
The second-order surface nonlinear problem, under \emph{undepleted-pump approximation}, can be described by the following driven wave equation with the nonlinear polarization of medium as a source term [18]

\begin{eqnarray}\label{eq1}
\nabla^2\v{E}^{(\nu)}+{k^{(\nu)}}^2\v{E}^{(\nu)}=-\frac{\nu^2}{\varepsilon_0c^2}\v{P}^{(\nu)}
\end{eqnarray}
where $k^{(\nu)}$ is the wave number at frequency $\nu$; $\v{P}^{(\nu)}$ is the surface nonlinear polarization at frequency $\nu$. It has the form $\v{P}^{(\nu)}=\varepsilon_0\overline{\overline\chi}^{(2)}:\v{E}_1^{(\omega_1)}\v{E}_2^{(\omega_2)}$ and $\nu=\omega_1+\omega_2$ for SFG. The SHG process is a special case of SFG with $\omega_1=\omega_2$ and $\nu=2\omega_1$.
$\overline{\overline\chi}^{(2)}$ is the surface second-order nonlinear susceptibility tensor. $\v{E}_1^{(\omega_1)}$ and $\v{E}_2^{(\omega_2)}$ denote the fundamental fields at frequencies $\omega_1$ and $\omega_2$, respectively. The symbol ``:" is a tensor operator functions as $P_i^{(\nu)}=\sum_{j,k}\chi_{ijk}^{(2)}E_{1j}^{(\omega_1)}E_{2k}^{(\omega_2)}$.
Metals are centrosymmetric materials. The surface of metal nanoparticles has an isotropic symmetry with a mirror plane perpendicular to it. The surface susceptibility tensor has only three non-vanishing and independent elements
\begin{eqnarray}\label{ChSHGeq33}
\v{P}^{(\nu)}=\varepsilon_0\hat{n}\Big[\chi_{\bot\bot\bot}^{(2)}\v{E}_{1\hat{n}}\v{E}_{2\hat{n}}
+\chi_{\bot\|\|}^{(2)}\big(\v{E}_{1\hat{t}_1}\v{E}_{2\hat{t}_1}+ \\ \nonumber 
\v{E}_{1\hat{t}_2}\v{E}_{2\hat{t}_2}\big)\Big] + \hat{t}_1\chi_{\|\bot\|}^{(2)}\v{E}_{1\hat{n}}\v{E}_{2\hat{t}_1}+\hat{t}_2\chi_{\|\bot\|}^{(2)}\v{E}_{1\hat{n}}\v{E}_{2\hat{t}_2}
\end{eqnarray}
where $\perp$ and $\parallel$ refer to the orthogonal and tangential components to the nanoparticle surface; $(\hat{t}_1,\hat{t}_2,\hat{n})$ is a system of three orthogonal vectors locally defined on the particle surface. The contribution of tangential and normal components of the surface nonlinear polarization are taken into account by the nonlinear surface electric and magnetic current sources
\begin{eqnarray}\label{ChSHGeq34}
\left\{ \begin{array}{ll}
\v{J}_0^{(\nu)} = -i\nu\v{P}_{t}^{(\nu)} \\
\v{M}_0^{(\nu)}= \frac{1}{\varepsilon^\prime}\hat{n}\times\nabla_S P_{n}^{(\nu)} \\
\end{array} \right.
\end{eqnarray}
where $\v{P}_{t}^{(\nu)}$ and $P_{n}^{(\nu)}$ are the tangential and normal components of $\v{P}^{\nu}$. The BEM method is applied to solve the driven wave equation by invoking the Love's equivalence principle.
The domain of the electromagnetic field is divided into the interior of the metal domain $V_i$, the exterior medium $V_e$ and the interface $S$. The object is illuminated by the plane wave source $\v{E}^{inc}$.  The equivalent currents $\v{J}_e^{(\nu)}, \v{M}_e^{(\nu)}$  positioned on the external page $S^{+}$ produce the scattered field in the region $V_e$  and null field in the region $V_i$ while the equivalent currents $\v{J}_i^{(\nu)}, \v{M}_i^{(\nu)}$, defined on the internal side $S^{-}$, produce the total field in the region $V_i$ and null field in the region $V_e$.
\begin{eqnarray}\label{ChSHGeq19}
\left. \begin{array}{ll}
\v{r}\in V_\ell, \, \v{E}_\ell^{(\nu)}(\v{r})\\
\v{r}\notin V_\ell,\,  0
\end{array} \right\}
=i\nu\mu_\ell\int_S\overline{\v{G}}^{(\nu)}\left(\mathbf{r},\mathbf{r}^\prime\right)\cdot\v{J}_\ell^{(\nu)}(\v{r}^\prime)d\v{r}^\prime \\ \nonumber
-\int_S\nabla\times\overline{\v{G}}^{(\nu)}\left(\mathbf{r},\mathbf{r}^\prime\right)\cdot\v{M}_\ell^{(\nu)}(\v{r}^\prime)d\v{r}^\prime
\end{eqnarray}
where $\overline{\v{G}}^{(\nu)}(\v{r},\v{r}^\prime)=(\overline{\v{I}}+{k^{(\nu)}}^{-2}\nabla\nabla)\exp{(ik^{(\nu)}R)}/4\pi R$ is the dyadic Green's function at frequency $\nu$. Here, $R=|\mathbf{r}-\mathbf{r}^\prime|$. $\ell=e,\,i$ denote the exterior and interior region of the object, respectively. The magnetic field has similar representation. The equivalent currents satisfy
\begin{eqnarray}\label{eq7}
\left\{ \begin{array}{cc}
\v{J}_i^{(\nu)}+\v{J}_e^{(\nu)}=\v{J}_0^{(\nu)} \\
\v{M}_i^{(\nu)}+\v{M}_e^{(\nu)}=\v{M}_0^{(\nu)}
\end{array} \right.
\end{eqnarray}

The surfaces of the nanostructures are discretized with triangular mesh. The equivalent currents are expanded with Rao-Wilton-Glisson (RWG) basis functions [19]. A matrix system is then constructed by exploiting the Galerkin testing procedure. A modified Poggio-Miller-Chang-Harrington-Wu-Tsai (PMCHWT) formulation [20] is used to ensure accurate solutions even at resonant conditions. The PMCHWT matrix equation can be written as
\begin{eqnarray}\label{eq9}
\overline{\v{C}}^{(\nu)}\cdot\v{X}^{(\nu)}=\v{Y}^{(\nu)},
\end{eqnarray}
The impedance matrix $\overline{\v{C}}^{(\nu)}$ is
\begin{eqnarray}\label{eq10}
\overline{\v{C}}^{(\nu)}= \left [\begin{array}{cccc}
\mathscr{L}_e^{(\nu)}& \mathscr{K}_e^{'(\nu)} & -\mathscr{L}_i^{(\nu)} & -\mathscr{K}_i^{'(\nu)} \\
-\mathscr{K}_e^{'(\nu)}& \eta_e^{-2}\mathscr{L}_e^{(\nu)} & \mathscr{K}_i^{'(\nu)} & -\eta_i^{-2}\mathscr{L}_i^{(\nu)} \\
\mathscr{I} & 0 & \mathscr{I} & 0 \\
0 &\mathscr{I} & 0 &  \mathscr{I} \nonumber \\
\end{array}\right ]
\end{eqnarray}
where $\mathscr{L}_\ell^{(\nu)}=i\nu\mu_\ell\overline{\v{G}}^{(\nu)}$ and $\mathscr{K}_\ell^{'(\nu)}$ is the principal value part of the $\mathscr{K}_\ell^{(\nu)}$ operator with $\mathscr{K}_\ell^{(\nu)}=-\nabla\times\overline{\v{G}}^{(\nu)}$. $\eta_\ell=\sqrt{\mu_\ell/\epsilon_\ell},$ with $\ell=e,\,i$.
The vector of unknowns $\v{X}^{(\nu)}$ and the excitation $\v{Y}^{(\nu)}$ are
\begin{eqnarray}\label{eq11}
\v{X}^{(\nu)}= \left [\begin{array}{c}
\v{J}_e^{(\nu)} \\
\v{M}_e^{(\nu)} \\
\v{J}_i^{(\nu)} \\
\v{M}_i^{(\nu)} \\
\end{array}\right ], \qquad
\v{Y}^{(\nu)}= \left [\begin{array}{c}
\frac{1}{2}\hat{n}\times\v{M}_0^{(\nu)} \\
-\frac{1}{2}\hat{n}\times\v{J}_0^{(\nu)} \\
\v{J}_0^{(\nu)} \\
\v{M}_0^{(\nu)} \nonumber \\
\end{array}\right ]
\end{eqnarray}

Equation \eqref{eq9} can be used to solve the fundamental fields as well by setting $\nu=\omega_\alpha$, with $\alpha=1,2$ for frequency $\omega_1$ and $\omega_2$, respectively. The driven source now is the incident excitation
\begin{eqnarray}\label{ChSHGeq31}
\qquad\left\{ \begin{array}{ll}
\v{J}_0^{(\omega_\alpha)}=\hat{n}\times\v{H}^{(\omega_\alpha, inc)}\\
\v{M}_0^{(\omega_\alpha)}=-\hat{n}\times\v{E}^{(\omega_\alpha, inc)}
\end{array} \right.
\end{eqnarray}

The BEM formulation can model the nonlinear scattering from arbitrarily shaped particles efficiently since it only requires surface discretization. In addition, measured material parameters can be used directly. In this work, only the normal component $\chi_{\perp\perp\perp}^{(2)}$ of the surface susceptibility tensor is considered, since it is the dominant term of the surface response of metallic nanoparticles. Note that other components are theoretically allowed, but they weakly contribute to the total SH response.

\section{Numerical Results}
\subsection{Verification of the Algorithm}
We first validate the BEM algorithm by comparing sum-frequency (SF) and second-harmonic (SH) scattering patterns of spherical nanoparticles with the nonlinear Mie solutions [21-22]. For SFG, two incident plane waves propagating in different angles with different frequencies are superposed as shown in Fig. 1. Fig. 2(a) shows the comparison of SF scattering patterns. Very good agreement is observed. Here, the radius of the sphere is $R=500\,\textrm{nm}$; the wavelengths of the incident waves are $\lambda_1=800\,\textrm{nm}$ and $\lambda_2=3,447\,\textrm{nm}$, respectively; the opening angle between the source waves is $\beta=15^\circ$; the index of refraction is set to unity for all wavelengths. The asymmetry in the SF scattering pattern is due to the nonzero opening angle between source beams. For SHG, a single electromagnetic field is taken as the source. A gold sphere with radius $R=50\,\textrm{nm}$ is excited by a plane wave, propagating along the positive direction of the $z$ axis and linearly polarized along $x$. The exciting wavelength $\lambda=520\,\textrm{nm}$ corresponds to the plasmon resonance of the gold spherical particle. The dielectric constant for gold is taken from experimental data [23]. We can see that the SH scattering pattern calculated by the BEM method agrees well with the nonlinear Mie theory result.

\begin{figure}[!h]
\begin{center}
\includegraphics[width=2.8in]{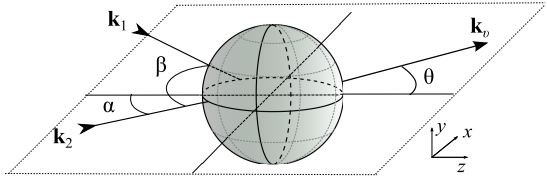}
  \caption{Overview of the relevant parameters in the model: the sum-frequency and source waves have wave vectors $\v{k}_{\nu}$, $\v{k}_1$ and $\v{k}_2$ in order of decreasing frequency. The angle between the propagation direction of the lowest frequency wave and the positive $z$ axis is $\alpha$; the opening angle between source waves is $\beta$. The sum-frequency scattering pattern is parameterized using the scattering angle $\theta$ [22].}
\label{fig1}
\end{center}
\end{figure}

\begin{figure}[!h]
\begin{center}
\includegraphics[width=3.5in]{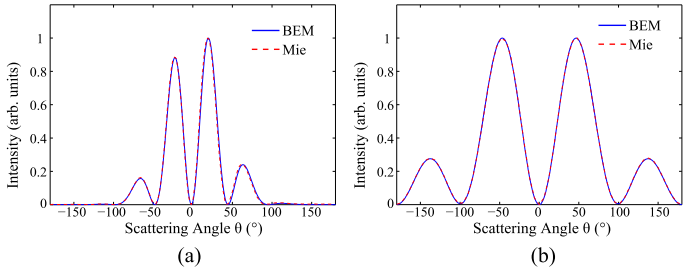}
  \caption{Comparisons of SF and SH scattering patterns with the nonlinear Mie solutions. (a) SF scattering pattern. The simulation parameters are $R=500\,\textrm{nm}$; $\lambda_1=800\,\textrm{nm}$ and $\lambda_2=3,447\,\textrm{nm}$; $\beta=15^{\circ}$. (b) SH scattering pattern. The simulation parameters are $R=50\,\textrm{nm}$; $\lambda_1=520\,\textrm{nm}$.}
\label{fig2}
\end{center}
\end{figure}

\subsection{SFG from Multi-resonant Nanoantenna}
The MR-NA consists of two gold arms of different lengths $L_1$ and $L_2$ ($L_1<L_2$). The separation between two arms is fixed to $20\,\textrm{nm}$. The width and height of the antenna arms are $40\,\textrm{nm}$. The lengths of the metal arms correspond to two resonant frequencies $\omega_1$ and $\omega_2$, respectively. Fig. 3 shows the radiation power as a function of the incident wavelength with different arm lengths. Due to the asymmetry of the antenna design, two resonant peaks are observed and they are red-shifted with the increase of the arm lengths. The insets in Fig. 3 show the distribution of fundamental frequency and SF equivalent electric currents on the surface of the MR-NA. The resonant peak at $\lambda_1=650\,\textrm{nm}$ corresponds to the half-wave resonance of the short arm as presented in inset (i). Similarly, the resonant peak at $\lambda_2=860\,\textrm{nm}$ corresponds to the long arm resonance as presented in inset (ii). The nonlinear equivalent currents at SF with $\lambda_\nu=372\,\textrm{nm}$ are concentrated near the gap and corners of the arms.

\begin{figure}[!h]
\begin{center}
\includegraphics[width=3.0in]{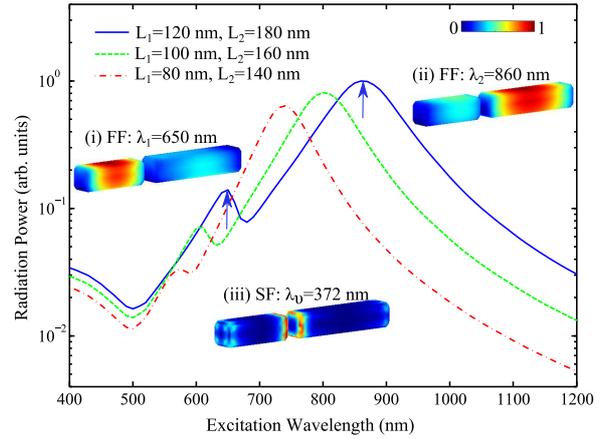}
  \caption{Radiation power of the MR-NA as a function of the incident wavelength with different arm lengths. The arrows denote the resonant peaks of the MR-NA with arm lengths $L_1=120\,\textrm{nm}$ and $L_2=180\,\textrm{nm}$. The inset shows the distribution of fundamental and SF equivalent electric currents on the surface of the MR-NA.}
\label{fig3}
\end{center}
\end{figure}

\subsection{SHG from Particle-in-cavity Nanoantenna}
The configuration of the investigated PIC-NA is depicted in Fig. 4(a). A gold nanosphere with a diameter $D=40\,\textrm{nm}$ is located inside a gold rounded-edge nanocup cavity separated by a small gap $g=5\,\textrm{nm}$. The nanocup cavity used is a truncated hemispherical nanoshell with external and inner radii being $R_1=120\,\textrm{nm}$ and $R_2=160\,\textrm{nm}$, respectively. Depending on the angle between the center of the nanosphere and the symmetry axis of the nanocup, the nanoantenna possesses symmetric ($\beta=0^\circ$) or asymmetric ($\beta=30^\circ$) geometry. The PIC-NA is illuminated by a $y$-polarized plane wave at the normal incidence from the top side. The linear and nonlinear responses of the PIC-NA are numerically simulated by the BEM method.

\begin{figure}[!h]
\begin{center}
\includegraphics[width=3.3in]{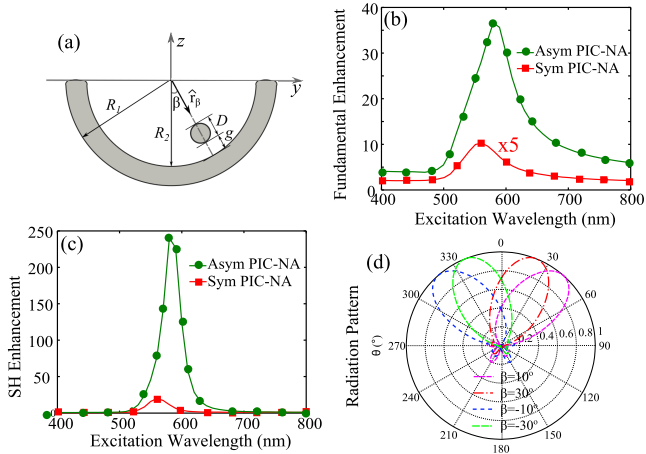}
  \caption{(a) Schematic of the PIC-NA ($yoz$ plane). A nanosphere (diameter $D$) is inside a nanocup cavity (external and inner radii being $R_1$ and $R_2$) separated by a gap $g$. The angle between the center of the nanosphere and the symmetry axis of the nanocup is $\beta$. (b) Fundamental and (c) SH enhancement spectra of the symmetric and asymmetric PIC-NA. (d) The steering of the main beam of the radiation pattern ($yoz$ plane) at the resonant wavelength by manipulating the position of particle.}
\label{fig4}
\end{center}
\end{figure}

Placing a metallic nanoparticle inside a cavity produces extremely strong field enhancements at the particle-cavity gap when one of the cavity modes is resonant with the cavity-dressed nanoparticle mode. The fundamental field enhancement is investigated at a fixed point (the center of the gap). The enhancement factor is defined as the ratio of the magnitude of the scattered field at the center of the gap to the magnitude of the incident field. Fig. 4(b) shows the calculated fundamental field enhancement spectra of the PIC-NA. The enhancement factors of roughly $2$ at $\lambda=560\,\textrm{nm}$ and $37$ at $\lambda=580\,\textrm{nm}$ are found for the symmetry and asymmetry cases, respectively. Because the SH field increases as the square of the fundamental field, a strong near-field at the fundamental frequency is particularly important for efficient SHG enhancement.

The SH enhancement factor is defined as the ratio of the SH intensity of the PIC-NA to the summation of the SH intensities of the single nanosphere and nanocup. Fig. 4(c) shows the SH enhancement factor. The correlation between fundamental field enhancement spectra (Fig. 4(b)) and the SH intensity spectra (Fig. 4(c)) demonstrates that the SHG from PIC-NA is boosted by the enhanced field intensity arising from the gap plasmonic mode. The radiation pattern of the PIC-NA is shown in Fig. 4 (d). Unidirectional radiation is observed. This important feature further enhances the SHG in the far field and facilitates the detection of the generated SH waves.  Moreover, beam-steering feature is achieved by changing the position of the nanosphere.

\section{Conclusion}
In summary, the boundary element method is developed for modeling surface nonlinear scattering from plasmonic nonlinear nanoantennas. The method is validated by comparing far-field SF and SH scattering patterns with Mie theory solutions. The SFG from the MR-NA, where two incident frequencies correspond to the fundamental resonances of the two arms, is analyzed by the developed method. The SHG from the PIC-NA is also modeled by the developed method. Unidirectional radiation of asymmetric PIC-NA is realized and the radiation direction can be controlled by the position of the nanoparticle. The directional beam steering offered by the proposed PIC-NA has promising applications such as nonlinear sensing, spectroscopy and frequency generation.

\section*{ACKNOWLEDGEMENT}
This work was supported in part by the Research Grants Council of Hong Kong (GRF 17207114 and GRF 17210815), NSFC 61271158, Hong Kong UGC AoE/P--04/08, and by the US NSF 1218552 and NSF 1609195.

\end{document}